\documentclass[twocolumn,showpacs]{revtex4}

\usepackage{graphicx}%
\usepackage{dcolumn}
\usepackage{amsmath}

\makeatletter
\def\btt#1{\texttt{\@backslashchar#1}}%
\DeclareRobustCommand\bblash{\btt{\@backslashchar}}%
\makeatother

\begin{document}

\preprint{HEP/123-qed}

\title[Short Title]{Terahertz radiation from magnetoresistive Pr$_{\text{0.7}}$Ca$_{\text{0.3}}$MnO$_{\text3}$ thin films}

\author{Noriaki Kida}
\email{kida@rcsuper.osaka-u.ac.jp} 
\affiliation{Research Center for Superconductor Photonics, Osaka University, 2-1 Yamadaoka, Suita, Osaka 565-0871, Japan}

\affiliation{CREST, Japan Science \& Technology Corporation (JST), 2-1 Yamadaoka, Suita, Osaka 565-0871, Japan}

\author{Masayoshi Tonouchi}

\affiliation{Research Center for Superconductor Photonics, Osaka University, 2-1 Yamadaoka, Suita, Osaka 565-0871, Japan}

\affiliation{CREST, Japan Science \& Technology Corporation (JST), 2-1 Yamadaoka, Suita, Osaka 565-0871, Japan}

\date{\today}

\begin{abstract}
Terahertz (THz) radiation with its spectrum extending up to 1 THz has been observed by an illumination of femtosecond optical pulses to optical switching devices fabricated on magnetoresistive manganite thin films; Pr$_{0.7}$Ca$_{0.3}$MnO$_3$. The THz radiation strongly depends on temperature $T$ and its $T$ trend reverses sign across charge-orbital and spin ordering $T$'s.

\end{abstract}



\maketitle



The progress of a THz-pulse generation and detection system using the photoconductive switch excited by femtosecond optical pulses, \cite{DHAuston} caused to general appreciations of this phenomenon and it rapidly gained ground as a result of many subsequent experimental and theoretical efforts. \cite{ABonvalet} The studies are oriented toward to the discovery of the THz radiation from various kind of materials including semiconductors, quantum wells, and superconductors. \cite{ABonvalet,PKBenicewicz,MHangyo,MTani} The basic mechanism of the THz radiation from these materials can be understood in the following classical Maxwell's equation, \cite{ABonvalet}
\begin{equation}
\mbox{\boldmath$E$}\propto\frac{\partial\mbox{\boldmath$J$}}{\partial t},
\label{eqn.1}
\end{equation}
where {\boldmath$E$} is the radiation field at the far-field and {\boldmath$J$} the current. In other words, the transient current, typically of the order of subpicosecond, is created by exciting carriers with ultrafast optical pulses, which in turn produce the THz-pulse propagating into the free space based on Eq. (\ref{eqn.1}). This manifests itself as the optical conversion via the transient photocurrent by the optical pulses.

In this letter, we show the finding of the THz radiation from magnetoresistive manganites and also describe how its characteristics vary with external stimulations. To the best of our knowledge, the THz radiation from magnetic materials has not been observed so far.

Hole-doped manganites with a perovskite structure, especially Pr$_{0.7}$Ca$_{0.3}$MnO$_3$ (PCMO) investigated here, has attracted great interests not only because of their potential applications to colossal magnetoresistive (CMR) devices but also their following unique features. \cite{YTokura} It exhibits insulating behavior with temperature $T$; charge and orbital ordering (a checkerboard pattern of Mn$^{3+}$ and Mn$^{4+}$ ions with 1:1 ratio) occur at $T_{\text{CO/OO}}$ and CE-type antiferromagnetic ordering develops below $T_{\text{N}}$. Below $T_{\text{CA}}$, it transforms to the charge-orbital ordered spin-canted antiferromagnetic insulator. \cite{YTomioka} This phase can be easily melted to the charge-orbital disordered ferromagnetic metallic state by external perturbations such as magnetic field (CMR effect), \cite{YTomioka} stress field, \cite{YMoritomo} electric field, \cite{AAsamitsu} other transition metal doping at Mn-site, \cite{BRaveau} electron beam irradiation, \cite{MHervieu} X-ray illumination, \cite{VKiryukhin} and visible-light illumination under an electric field \cite{KMiyano,MFiebig1,MFiebig2,MFiebig3,MFiebig4} or a magnetic field. \cite{YOkimoto1} Such dramatic phase controls of PCMO make them quite useful candidate for future applications. \cite{YTokura} Moreover, recent spectroscopic studies using pump-and-probe technique revealed that the fast optical response begins at a characteristic time of subpicosecond. \cite{MFiebig4} According to Eq. (\ref{eqn.1}), it is expected that the THz-pulse can also be generated from PCMO, however its mechanism is basically different from semiconductors and the ultrafast optical modulation of the charge, spin, and orbital ordering may be responsible for this effect. In this purpose, we found the THz radiation from PCMO thin films excited by femtosecond optical pulses.

PCMO thin films on MgO(100) substrates were prepared by pulsed laser deposition technique. Using conventional photolithographic and liftoff techniques, we formed the optical switching structure coupled to a bow-tie antenna fabricated on PCMO thin film. It consists of 3 mm long and 200 $\mu$m wide gold lines with a gap of 5 $\mu$m and a bow-angle of 60$^\circ$.

The THz-pulse generation and detection system is as follows. A mode-locked Ti:sapphire laser was used to generate 80 femtosecond optical pulses at the wavelength of 800 nm. The energy of the optical pulse exceeds the charge-gap energy $\sim$0.5 eV of PCMO and nearly corresponds to the charge transfer energy from O $2p$ to Mn $3d$ band. \cite{YOkimoto2} The pump-pulses chopped mechanically at 2 kHz were focused into the region of 30 $\mu$m in diameter by an object lens on the gap of the bow-tie antenna. In order to enhance the collection efficiency of the THz-pulse, we attached an MgO hemispherical lens on the backside of the MgO substrate. \cite{MTonouchi} The propagated THz-pulse was collimated and focused by a pair of off-axis paraboloidal mirrors. The THz detector was a bow-tie antenna made of Au/Ge/Ni alloy lines fabricated on the low-temperature grown GaAs thin film.

\begin{figure}[tbp]
\includegraphics[trim=100 163 100 115, clip, keepaspectratio=true, width=7cm]{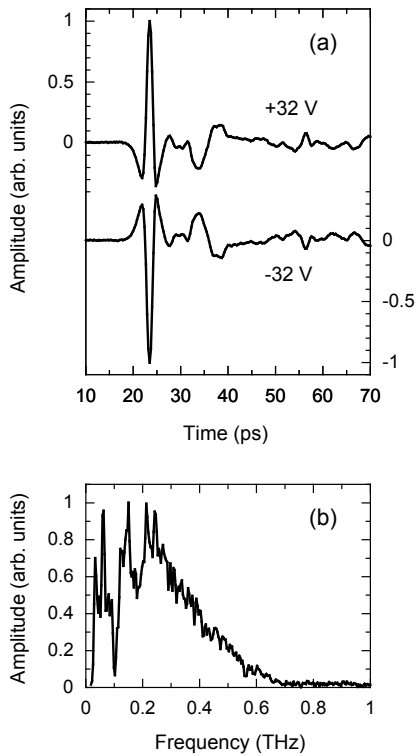}

\caption{(a) The measured THz-pulse in the time-domain of Pr$_{0.7}$Ca$_{0.3}$MnO$_3$ thin film at 23 K under the applied voltage of $\pm$32 V and the pump-pulse energy of 184 mW. (b) Amplitude spectrum in the frequency-domain obtained by the Fourier transformation of (a). Data is normalized with the value of the absolute maximum amplitude.}
\label{fig1}
\end{figure}

\begin{figure}[bp]
\includegraphics[trim=40 195 20 200, clip, keepaspectratio=true, width=7.5cm]{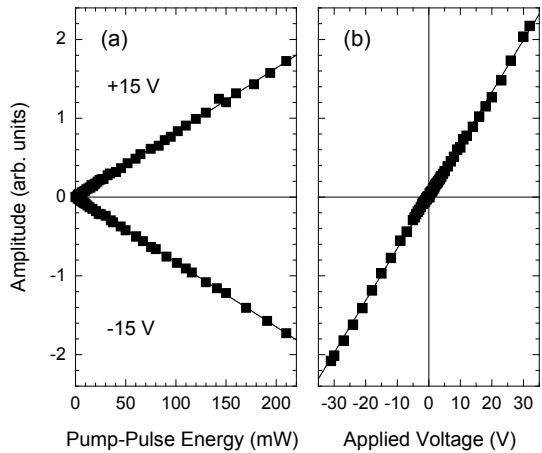}
\caption{(a) The pump-pulse energy at fixed applied voltages of $\pm$15 V and (b) the applied voltage at the fixed pump-pulse energy of 122 mW, dependence of the main peak amplitude $E_{\text{THz}}$ in the time-domain waveform, measured at 23 K. Data is normalized with the value of $E_{\text{THz}}$ under the applied voltage of 15 V and the pump-pulse energy of 122 mW.}
\label{fig2}
\end{figure}

We observed the THz radiation from PCMO thin film at 23 K under applied voltages of $\pm$32 V and the pump-pulse energy of 184 mW; the generated electromagnetic waveforms under above conditions are shown in Fig. \ref{fig1}(a). The main peaks were observed at around 24 ps (zero position in the time-delay is in arbitrary units). The pulsewidth (full-width at half maximum of the main peak) is estimated to be 1 ps using a Gaussian profile to fit the data. The change in the polarity of the applied voltage caused the reversal of the generated electromagnetic waveform. It was verified that both the applied voltage and the illumination of the femtosecond optical pulses are necessary to generate the THz-pulse.

These phenomena are not due to the local heating effect by the strong field caused by both the electric field and the femtosecond optical pulses, because the main peak amplitude as a function of $T$ changes around electronic and magnetic transition $T$'s as will be shown later.

The time-domain waveform in Fig. \ref{fig1}(a) were Fourier transformed to the frequency-domain; the obtained amplitude spectrum is shown in Fig. \ref{fig1}(b). It extends to $\sim$1 THz centered at around 0.2 THz. The dips around 0.1 and 0.18 THz can be seen, however the origin of these structures is not clear at present. The spectrum also shows an oscillating structure arising from the effect of multiple reflections of MgO substrate.

To get detailed THz radiation characteristics, we display in Figs. \ref{fig2}(a) and \ref{fig2}(b), the main peak amplitude $E_{\text{THz}}$ in the time-domain waveform as a function of the pump-pulse energy and the applied voltage, respectively. Solid lines show least-squares fits of the data (solid squares), assuming the linear response for external stimulations. $E_{\text{THz}}$ was found to be a linear function of the pump-pulse energy at fixed applied voltages of $\pm$15 V [Fig. \ref{fig2}(a)]. We also found that $E_{\text{THz}}$ depends linearly on the applied voltage at the fixed pump-pulse energy of 122 mW [Fig. \ref{fig2}(b)].


Our experimental results shown in Figs. \ref{fig1}(a) and \ref{fig2} suggest that the THz radiation from PCMO can be described by the current surge model, \cite{PKBenicewicz} which has been successfully applied to quantitatively explain THz radiation characteristics from a large \cite{PKBenicewicz} as well as small \cite{MTani} gap semiconductor photoconductive antenna used for this work.



In our experimental configuration, it is possible to induce the persistent insulator-metal (IM) transition by the visible light illumination under an electric field. \cite{KMiyano,MFiebig1,MFiebig2,MFiebig3,MFiebig4} When the THz-pulse generates and propagates into free space, we supposed that the persistent IM transition does not occur and which is confirmed by following experimental observations. First, we did not observe the threshold behavior for the THz radiation, which is a typical feature of the photo-induced transition  \cite{KMiyano,SKoshihara} as shown in Figs. \ref{fig2}(a) and \ref{fig2}(b). Second, we did not detect the breakdown of the applied voltage during our measurements as we were simultaneously monitoring the applied voltage by another voltmeter connected in the parallel configuration with the sample.

\begin{figure}[tbp]
\includegraphics[trim=100 280 150 200, clip, keepaspectratio=true, width=6cm]{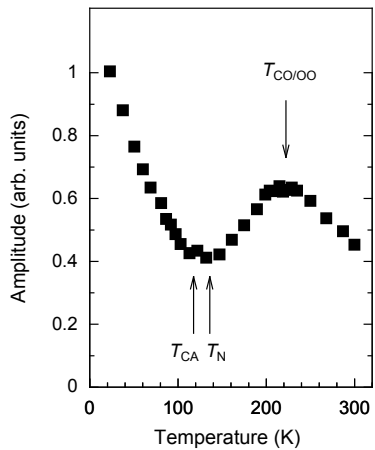}
\caption{Temperature dependence of the main peak amplitude $E_{\text{THz}}$ in the time-domain waveform at the fixed applied voltage of 15 V and the pump-pulse energy of 100 mW. Data is normalized with the value of $E_{\text{THz}}$ at 23 K.}
\label{fig3}
\end{figure}


Figure \ref{fig3} shows $T$ dependence of $E_{\text{THz}}$ at the applied voltage of 15 V and the pump-pulse energy of 100 mW on heating runs. During heating process, the position of the sample moves due to both the thermal expansion of the materials and the vibration of the cryostat. Therefore, we fixed a time-delay at the main peak and subsequently scanned the sample at each $T$ and the laser spot was tuned to the center of the gap before taking data. As clearly seen, $E_{\text{THz}}$ shows an interesting and strong $T$ dependence. The $T$ variation of the absorption coefficient of this material is negligible at the center of the laser-wavelength used in the photoexcitation. \cite{YOkimoto2} We found that $E_{\text{THz}}$ tends to decrease as $T$ is increased around 120 K. For further increase with $T$, $E_{\text{THz}}$ monotonically increases up to 230 K. Even at the room $T$, the THz radiation is clearly observed. These critical points matches with previously reported values; $T_{\text{CA}}$, $T_{\text{N}}$, and $T_{\text{CO/OO}}$. \cite{YTomioka} Moreover, the slope of $E_{\text{THz}}(T)$ changes in reverse around these $T$'s.




Recently, Fiebig {\it et al}. have reported that, the transient reflectivity change in single crystals of PCMO occurs rapidly within 1 ps, using the femtosecond pump-and-probe spectroscopic technique. \cite{MFiebig4} They interpreted this fast optical response as the photo-induced melting of the charge ordered insulating state based on their systematic optical studies. \cite{MFiebig2,MFiebig3,MFiebig4} In our preliminary results, not shown here, $E_{\text{THz}}$ transiently and nonlinearly decreases in intensity with increasing the applied voltage under the constant illumination of the femtosecond optical pulses, which may indicate the formation of the conducting path (the occurrence of the persistent photo-induced IM transition described before). Therefore, we surmise that the ultrafast formation of the photogenerated metallic patch embedded with the insulating matrix is essential to generate the THz-pulse and their nonequilibrium dynamics play an important role in understanding the observed $E_{\text{THz}}(T)$ behavior. Further studies using a dipole antenna are needed to clarify the generation mechanism and now in progress.

Our findings of the THz radiation from PCMO having a perovskite structure open the wide range of the related materials for future ultrafast optical devices. PCMO as well as other perovskite-type transition metal oxides may be potential candidates for intense THz emitters; the reflectivity of these materials at 800 nm used in the photoexcitation is about 20\%, so the light is effectively absorbed. Further experiment under a magnetic field is of great interest. From a fundamental viewpoint, the studies of the ultrafast phenomena in manganites, characterizations of the THz radiation in the time-domain are expected to give fruitful information in the underlying physics of manganites.

We thank Dr. M. Fiebig for giving fruitful comments and Dr. M. Misra and A. Quema for reading the manuscript.


\begin{references}

\bibitem{DHAuston} D. H. Auston, K. P. Cheung, and P. R. Smith, Appl. Phys. Lett. {\bf 45}, 284 (1984).
\bibitem{ABonvalet} A. Bonvalet and M. Joffre, in {\it Femtosecond Laser Pulses: Principles and Experiments}, edited by C. Rulli$\grave{\text e}$re (Springer-Verlag, Heidelberg, 1998) p. 285.
\bibitem{PKBenicewicz} P. K. Benicewicz, J. P. Roberts, and A. J. Taylor, J. Opt. Soc. Am. B {\bf 11}, 2533 (1994).
\bibitem{MHangyo} M. Hangyo, S. Tomozawa, Y. Murakami, M. Tonouchi, M. Tani, Z. Wang, K. Sakai, and S. Nakashima, Appl. Phys. Lett. {\bf 69}, 2122 (1996).
\bibitem{MTani} M. Tani, S. Matsuura, K. Sakai, and S. Nakashima, Appl. Opt. {\bf 36}, 7853 (1997).



\bibitem{YTokura} Y. Tokura, JSAP International No. 2, 12 (2000).
\bibitem{YTomioka} Y. Tomioka, A. Asamitsu, H. Kuwahara, Y. Moritomo, and Y. Tokura, Phys. Rev. B {\bf 53}, R1689 (1996).
\bibitem{YMoritomo} Y. Moritomo, H. Kuwahara, Y. Tomioka, and Y. Tokura, Phys. Rev. B {\bf 55}, 7549 (1997).
\bibitem{AAsamitsu} A. Asamitsu, Y. Tomioka, H. Kuwahara, and Y. Tokura, Nature (London) {\bf 388}, 50 (1997).
\bibitem{BRaveau} B. Raveau, A. Maignan, and C. Martin, J. Solid State Chem. {\bf 130}, 162 (1997).
\bibitem{MHervieu} M. Hervieu, A. Barnab$\acute{\text e}$, C. Martin, A. Maignan, and B. Raveau, Phys. Rev. B {\bf 60}, R726 (1999).
\bibitem{VKiryukhin} V. Kiryukhin, D. Casa, J. P. Hill, B. Keimer, A. Vigliante, Y. Tomioka, and Y. Tokura, Nature (London) {\bf 386}, 813 (1997).
\bibitem{KMiyano} K. Miyano, T. Tanaka, Y. Tomioka, and Y. Tokura, Phys. Rev. Lett. {\bf 78}, 4257 (1997).
\bibitem{MFiebig1} M. Fiebig, K. Miyano, Y. Tomioka, and Y. Tokura, Science {\bf 280}, 1925 (1998).
\bibitem{MFiebig2} M. Fiebig, K. Miyano, Y. Tomioka, and Y. Tokura, Appl. Phys. Lett. {\bf 74}, 2310 (1999).
\bibitem{MFiebig3} M. Fiebig, K. Miyano, T. Satoh, Y. Tomioka, and Y. Tokura, Phys. Rev. B {\bf 60}, 7944 (1999).
\bibitem{MFiebig4} M. Fiebig, K. Miyano, Y. Tomioka, and Y. Tokura, Appl. Phys. B {\bf 71}, 211 (2000).
\bibitem{YOkimoto1} Y. Okimoto, Y. Tokura, Y. Tomioka, Y. Onose, Y. Otsuka, and K. Miyano, Mol. Cryst. Liq. Cryst. {\bf 315}, 257 (1998).
\bibitem{YOkimoto2} Y. Okimoto, Y. Tomioka, Y. Onose, Y. Otsuka, and Y. Tokura, Phys. Rev. B {\bf 57}, R9377 (1998).
\bibitem{MTonouchi} M. Tonouchi, M. Tani, Z. Wang, K. Sakai, M. Hangyo, N. Wada, and Y. Murakami, IEEE Trans. Appl. Supercond. {\bf 7}, 2913 (1997).
\bibitem{SKoshihara} S. Koshihara, J. Lumin. {\bf 87-89}, 77 (2000).


\end{references}
\end{document}